\documentclass[aps,prd,onecolumn,superscriptaddress,floats,amsfonts,amsmath,amssymb,12pt]{revtex4-1}
\usepackage{graphicx}
\usepackage{epstopdf}
\usepackage{bm}
\usepackage[]{latexsym}

\usepackage{color}

\usepackage{caption2}

\usepackage{float}

\usepackage{lineno}

\begin{document}

\restylefloat{figure}

\title{Constraining the Mass Scale of a Lorentz-Violating Hamiltonian with the Measurement of Astrophysical Neutrino-Flavor Composition}

\author{Kwang-Chang Lai}
\email{kcl@mail.cgu.edu.tw}
\affiliation{Center for General Education, Chang Gung University, Kwei-Shan, Taoyuan, 333, Taiwan}

\author{Wei-Hao Lai}
\email{s9927525.py99g@g2.nctu.edu.tw}
\affiliation{Institute of Physics, National Chiao Tung University, Hsinchu, 300, Taiwan}

\author{Guey-Lin Lin}
\email{glin@mail.nctu.edu.tw}
\affiliation{Institute of Physics, National Chiao Tung University, Hsinchu, 300, Taiwan}

\begin{abstract}

We study Lorentz violation effects on flavor transitions of high energy astrophysical neutrinos. It is shown that the appearance of Lorentz violating  Hamiltonian can drastically change the flavor transition probabilities of 
astrophysical neutrinos.  Predictions of Lorentz violation effects on flavor compositions of astrophysical neutrinos arriving on Earth are compared with IceCube flavor composition measurement which analyzes astrophysical neutrino events in the energy range between $25~{\rm TeV}$ and $2.8~{\rm PeV}$. Such a comparison indicates that the future IceCube-Gen2 will be able to place stringent constraints on Lorentz violating Hamiltonian in the neutrino sector.   
We work out the expected sensitivities by IceCube-Gen2 on dimension-$3$ CPT-odd and dimension-$4$ CPT-even operators in Lorentz violating Hamiltonian. The expected sensitivities can improve on the current constraints obtained from other types of experiments by more than two orders of magnitudes for certain range of the parameter space.

\vspace{3mm}


\end{abstract}

\maketitle

\section{Introduction}
Although physical laws are believed to be invariant under Lorentz  transformation, violations of Lorentz symmetry might arise in string theory as discussed in~\cite{Kostelecky:1988zi,Kostelecky:1991ak}. It is possible to incorporate Lorentz violation (LV) effects in an observer-independent effective field theory, the so-called Standard-Model Extension (SME)~\cite{Colladay:1996iz,Colladay:1998fq}, which encompasses all the features of standard model particle physics and general relativity plus all possible LV operators~\cite{Kostelecky:2003fs,AmelinoCamelia:2005qa,Bluhm:2005uj}.  While LV signatures are suppressed by the ratio $\Lambda_{\rm EW}/m_{\rm P}$ with $\Lambda_{\rm EW}$ the electroweak energy scale and $m_{\rm P}$ the Planck scale, experimental techniques have been developed for probing such signatures~\cite{Kostelecky:2008ts,Mattingly:2005re}. The effects of LV on neutrino oscillations were pointed out in~\cite{Kostelecky:2003cr,Kostelecky:2003xn,Kostelecky:2004hg}. One can categorize LV effects to neutrino flavor transitions into three aspects: the modifications to energy dependencies of neutrino oscillation probabilities, the directional dependencies of oscillation probabilities, and the modifications to neutrino mixing angles and phases. In the standard vacuum oscillations of neutrinos, the oscillatory behavior of flavor transition probability is determined by the dimensionless variable $\Delta m^2 L/E$ with $\Delta m^2$  the neutrino mass-squared difference, $L$ the neutrino propagation distance, and $E$ the neutrino energy.  
This dependence results from the Hamiltonian $H_{\rm SM}=UM^2U^{\dagger}/2E$ with $M^2_{ij}=\delta_{ij}(m_j^2-m_1^2)$. The extra terms in Lorentz violating Hamiltonian $H_{\rm LV}$ introduces $L$ and $LE$ dependencies into the oscillation probability, in addition to the standard $L/E$ dependence. The directional dependence of oscillation probability is due to the violation of rotation symmetry in $H_{\rm LV}$. The coefficients of LV operators change periodically as the Earth rotates daily about its axis. This induces temporal variations of neutrino oscillation probability at multiples of sidereal frequency $\omega_\oplus\approx 2\pi/({\rm 23 \ h \ 56 \ min})$. Finally the full Hamiltonian $H \equiv H_{\rm SM}+H_{\rm LV}$ is diagonalized by the unitary matrix $V$ which differs from $U$ due to the appearance of $H_{\rm LV}$. Hence the values of neutrino mixing angles and phases associated with $V$ deviate from those associated with $U$. Such deviations increase with neutrino energies since $H_{\rm SM}$ is $\mathcal{O}(E^{-1})$ while $H_{\rm LV}$ contains $\mathcal{O}(E^0)$ and
$\mathcal{O}(E)$ terms. 

Experimentally, effects of Lorentz violation on neutrino oscillations have been investigated in short-baseline neutrino beams \cite{Auerbach:2005tq,Adamson:2008aa,AguilarArevalo:2011yi,Adamson:2012hp}, in long-baseline neutrino beams \cite{Adamson:2010rn,Rebel:2013vc}, in reactor neutrinos at Double Chooz \cite{Abe:2012gw,Diaz:2013iba}, and in atmospheric neutrinos at IceCube \cite{Abbasi:2010kx} and Super-Kamiokande \cite{Abe:2014wla}. These experiments probe either the spectral anomalies of the oscillated neutrino flux or the sidereal variations of neutrino oscillation probabilities. In this paper, we shall focus on LV effects to neutrino mixing angles and phases. As mentioned before, these effects grow with neutrino energies. Thus it is ideal to probe such effects through the flavor transitions of high energy astrophysical neutrinos~\cite{dispersion}. For simplicity, we only consider isotropic LV effects. 

The observation of  high energy astrophysical neutrinos by IceCube \cite{Aartsen:2013bka,Aartsen:2013jdh,Aartsen:2013eka,Aartsen:2014gkd} is a significant progress in neutrino astronomy and provides new possibilities for testing neutrino properties. The first result by IceCube on the flavor composition of observed 
astrophysical neutrinos has been published in \cite{Aartsen:2015ivb}, and was updated in \cite{Aartsen:2015knd} by a combined-likelihood analysis taking into account more statistics. Meanwhile, independent efforts have been made to determine neutrino flavor compositions from IceCube data \cite{Mena:2014sja,Winter:2014pya,Chen:2014gxa,Palomares-Ruiz:2015mka,Palladino:2015zua}.  
As we shall see in latter sections, the flavor measurement in~\cite{Aartsen:2015knd} is not yet able to constrain $H_{\rm LV}$ more stringently than the previous experiments. Fortunately there is an active plan for extending the current IceCube detector to a larger volume, which is referred to as 
IceCube-Gen2~\cite{Aartsen:2014njl,Aartsen:2015dkp}.  This extension shall increase the effective area of the current 86-string detector up to a factor of 5. The expected improvement on neutrino flavor discrimination by IceCube-Gen2 has been studied in~\cite{Shoemaker:2015qul}.  Using this result, we shall study sensitivities of IceCube-Gen2 to the parameters of $H_{\rm LV}$.    

Astrophysical neutrinos are commonly produced by either $pp$ or $p\gamma$ collisions at astrophysical sources. For sufficiently high energies, $pp$ collisions produce 
equal number of $\pi^+$ and $\pi^-$, which decay to neutrinos through $\pi^+\rightarrow\mu^+ +\nu_\mu\rightarrow e^+ +\nu_\mu+\nu_e +\bar{\nu}_\mu$ and $\pi^- \rightarrow\mu^- +\bar{\nu}_\mu\rightarrow e^- +\bar{\nu}_\mu+\bar{\nu}_e +\nu_\mu$. This leads approximately to the flux ratio $\Phi^0(\nu_e):\Phi^0(\nu_\mu):\Phi^0(\nu_\tau)=1/3:2/3:0$ for both neutrinos and anti-neutrinos. Here $\Phi^0(\nu_\alpha)$ denotes generically 
the flux of neutrino or anti-neutrino of flavor $\alpha$. This type of source is referred to as the pion source.
A more detailed study on the neutrino flavor fraction with the consideration of  neutrino spectral index is given in \cite{Lipari:2007su}. For an $E^{-2}$ spectrum, the neutrino flavor fraction at the source is $(f^0_e,f^0_{\mu},f^0_{\tau})=(0.35,0.65,0)$, where $f^0_{\alpha}\equiv \Phi^0(\nu_\alpha)/(\Phi^0(\nu_e)+\Phi^0(\nu_\mu)+\Phi^0(\nu_\tau))$. However, for the purpose of this work, it suffices to take  $(f^0_e,f^0_{\mu},f^0_{\tau})=(1/3,~2/3,~0)$.
We note that the secondary muons in some astrophysical objects can lose energy quickly by synchrotron cooling in magnetic fields or interactions with matter before their decays.  Hence the neutrino flavor fraction at the source   
becomes $(0,~1,~0)$. This type of source is referred to as the muon-damped source \cite{Kashti:2005qa,Kachelriess:2007tr,Hummer:2010ai}. In fact, there are also cases that the flavor fraction of astrophysical neutrinos at the source is energy dependent. For example, the flavor fraction of neutrinos can gradually changes from $(1/3,~2/3,~0)$ at lower energies to $(0,~1,~0)$ at high energies. Such a phenomenon has been discussed in~\cite{Kashti:2005qa,Kachelriess:2007tr} and investigated systematically in~\cite{Hummer:2010ai}. The latter work also discusses sources with flavor fractions different from those of the pion source and muon-damped source. While a general study should consider the energy dependence of neutrino flavor fraction and variations of neutrino flavor fractions among different sources, we shall only focus on the simplified scenario that all sources of astrophysical neutrinos arising from $pp$ collisions possess an energy independent flavor fraction for neutrinos at $(1/3,~2/3,~0)$.       

The production mechanism of astrophysical neutrinos 
with $p\gamma$ collisions is more complicated. The leading process of this category is $p\gamma\to n\pi^+$ which gives rise to the flavor fraction $(1/2,~1/2,~0)$ for neutrinos and $(0,~1,~0)$ for anti-neutrinos.
The sub-leading process is $p\gamma \to p\pi^+\pi^-$ which is non-negligible when the spectral index $\beta$ of the target photon is harder than $1$ \cite{Murase:2005hy,Baerwald:2010fk}. This process produces equal number of neutrinos and anti-neutrinos 
with a common flavor fraction $(1/3,~2/3,~0)$. Since the flavor fraction of neutrinos produced by $p\gamma$ collisions is relatively uncertain, we will not consider astrophysical neutrinos produced by such a mechanism.

We note that effects of new-physics Hamiltonian (with Lorentz violation as a special case), parametrized as $(E_{\nu}/\Lambda_n)^n U_nO_nU_n^{\dagger}$,
on the flavor transitions of astrophysical neutrinos were discussed in~\cite{Arguelles:2015dca,Katori:2016eni} for $n=0$ and $1$ (similar discussions were also given in ~\cite{Barenboim:2003jm,Hooper:2005jp,Bustamante:2010nq,Bustamante:2015waa}), and comparisons with earlier IceCube flavor measurement~\cite{Aartsen:2015ivb} were made. The authors scan all possible structures of the mixing matrix $U_n$ for given new-physics scales $\Lambda_n$ and $O_n$ and determine the allowed range of astrophysical neutrino flavor fractions on Earth resulting from the full Hamiltonian $H=H_{\rm SM}+(E_{\nu}/\Lambda_n)^n U_nO_nU_n^{\dagger}$.  In our work, we shall focus on LV effects 
which are parameterized in a different form from the above new-physics Hamiltonian. We shall discuss current and future constraints on LV effects by comparing the predicted neutrino flavor fraction with the range of flavor fraction 
measured by the current IceCube detector ~\cite{Aartsen:2015knd} and that expected~\cite{Shoemaker:2015qul} in the future IceCube-Gen2 detector. Our results can be directly compared with the previously most stringent constraints obtained by Super-Kamiokande~\cite{Abe:2014wla}.   
 
This paper is organized as follows. In Sec. II, we incorporate LV effects into the full neutrino Hamiltonian in the framework of SME. We then study analytically 
the flavor transition of astrophysical neutrinos assuming the dominance of $H_{\rm LV}$ over $H_{\rm SM}$. As stated before such a dominance is possible for high energy astrophysical neutrinos. 
We discuss constraints on LV effects by the current IceCube flavor measurement. Such discussions pave the way for detailed numerical studies in the next section.   In Sec. III, we study the flavor transitions of astrophysical neutrinos with the 
full Hamiltonian $H=H_{\rm SM}+H_{\rm LV}$. The expected sensitivities of IceCube-Gen2  to $H_{\rm LV}$ are studied.  We conclude in Sec. IV.

\section{Lorentz Violation in Neutrino Oscillations}

LV effects in neutrino oscillations are incorporated by introducing an additional Lorentz violating term $H_{\rm LV}$ to the full Hamiltonian of the neutrino. Hence
\begin{equation}
H=H_{\rm SM} + H_{\rm LV}, \label{HLV}
\end{equation}
where $H_{\rm SM}\equiv UM^2U^{\dag}/2E$ is the standard model neutrino Hamiltonian in vacuum with $M^2$  the neutrino mass matrix
\begin{equation}
M^2 =\left(
                            \begin{array}{ccc}
                            0 &             0             & 0 \\
                            0 & \Delta m^2_{21} & 0 \\
                            0 &             0             & \Delta m^2_{31}
                            \end{array}
                            \right),
\end{equation}
and $U$ the PMNS matrix.
Here we do not consider matter effects due to neutrino propagations inside the Earth.  This is because we only focus on neutrino events with energies higher than few tens of TeV. In this case
the Earth regeneration effect to the neutrino flavor transition is negligible.  
For neutrinos, the general form of LV Hamiltonian is given by 
\begin{equation}
H_{\rm LV}^{\nu} =\frac{p_{\lambda}}{E} \left(
                            \begin{array}{ccc}
                            a_{ee}^{\lambda}                 &     a_{e\mu}^{\lambda}             & a_{e\tau}^{\lambda} \\
                            a_{e\mu}^{\lambda *} &      a_{\mu\mu}^{\lambda}                           & a_{\mu\tau}^{\lambda } \\
                            a_{e\tau}^{\lambda *} &      a_{\mu\tau}^{\lambda *}       & a_{\tau\tau}^{\lambda}  
                            \end{array}
                            \right)
                   -\frac{p^{\rho}p^{\lambda}}{E} \left(
                            \begin{array}{ccc}
                                  c_{ee}^{\rho\lambda}             &     c_{e\mu}^{\rho\lambda}             & c_{e\tau}^{\rho\lambda}  \\
                            c_{e\mu}^{\rho\lambda *}  &       c_{\mu\mu}^{\rho\lambda}                          & c_{\mu\tau}^{\rho\lambda}  \\
                            c_{e\tau}^{\rho\lambda *} &      c_{\mu\tau}^{\rho\lambda *}        &  c_{\tau\tau}^{\rho\lambda} 
                            \end{array}
                            \right).
\end{equation} 
Since we shall only consider isotropic LV effects, we have the simplified form for $H_{\rm LV}^{\nu}$ given by~\cite{Kostelecky:2003cr}
\begin{equation}
H_{\rm LV}^{\nu} = \left(
                            \begin{array}{ccc}
                            a_{ee}^{T}                 &     a_{e\mu}^{T}             & a_{e\tau}^{T} \\
                            a_{e\mu}^{T*} &      a_{\mu\mu}^{T}                           & a_{\mu\tau}^{T } \\
                            a_{e\tau}^{T *} &      a_{\mu\tau}^{T *}       & a_{\tau\tau}^{T}  
                            \end{array}
                            \right)
                   - \frac{4E}{3} \left(
                            \begin{array}{ccc}
                                  c_{ee}^{TT}             &     c_{e\mu}^{TT}             & c_{e\tau}^{TT}  \\
                            c_{e\mu}^{TT *}  &       c_{\mu\mu}^{TT}                          & c_{\mu\tau}^{TT}  \\
                            c_{e\tau}^{TT*} &      c_{\mu\tau}^{TT *}        &  c_{\tau\tau}^{TT} 
                            \end{array}
                            \right),
\label{hamilton_nu}
\end{equation}
where $T$ is the time component of Sun-centered celestial equatorial
coordinate $(T,~X,~Y,~Z)$.
For anti-neutrinos, we have
\begin{equation}
H_{\rm LV} ^{\bar{\nu}}=-\left(
                            \begin{array}{ccc}
                            a_{ee}^{T}                 &     a_{e\mu}^{T}             & a_{e\tau}^{T} \\
                            a_{e\mu}^{T*} &      a_{\mu\mu}^{T}                           & a_{\mu\tau}^{T } \\
                            a_{e\tau}^{T *} &      a_{\mu\tau}^{T *}       & a_{\tau\tau}^{T}  
                            \end{array}
                            \right)^*
                   - \frac{4E}{3} \left(
                            \begin{array}{ccc}
                                  c_{ee}^{TT}             &     c_{e\mu}^{TT}             & c_{e\tau}^{TT}  \\
                            c_{e\mu}^{TT *}  &       c_{\mu\mu}^{TT}                          & c_{\mu\tau}^{TT}  \\
                            c_{e\tau}^{TT*} &      c_{\mu\tau}^{TT *}        &  c_{\tau\tau}^{TT} 
                            \end{array}
                            \right)^*.
\label{hamilton_nubar}
\end{equation}

The two terms on the right hand side of $H_{\rm LV}^{\nu,\bar{\nu}}$ are distinguished by their CPT transformation properties and dimensionality of the operators they are originated from.
The first term is CPT-odd and originated from dimension-$3$ operator while the second term is CPT-even and  originated from dimension-$4$ operator.
Diagonalizing the full Hamiltonian in Eq.~(\ref{HLV}) yields a new mass-flavor mixing matrix $V$. The neutrino flavor transition probability $P_{\alpha\beta} \equiv P(\nu_{\beta}\to \nu_{\alpha})$ is then given by
\begin{equation}
P_{\alpha\beta} = \delta_{\alpha\beta} - 4\sum_{j>i}\Re(V_{\beta j}V^*_{\beta i}V^*_{\alpha j}V_{\alpha i})\sin^2(L\Delta E_{ji}/2)
                          + 2\sum_{j>i}\Im(V_{\beta j}V^*_{\beta i}V^*_{\alpha j}V_{\alpha i})\sin^2(L\Delta E_{ji}),
\end{equation} 
where $\Delta E_{ji}\equiv E_j - E_i$ is the difference between the energy eigenvalues. For high-energy astrophysical neutrinos, $L$ is so large that the rapid oscillating terms are averaged out so that
\begin{equation}
P_{\alpha\beta} = \sum^3_{i=1}|V_{\alpha i}|^2|V_{\beta i}|^2 \label{prob}.
\end{equation} 
Since $P_{\alpha\beta}$ depends only on the elements of $V$, the neutrino flavor composition observed on the Earth for a given astrophysical neutrino source is affected by LV parameters. Therefore, the measurement of neutrino flavor fraction by neutrino telescopes such as IceCube is useful for constraining
LV parameters. For convenience in discussions, we shall first concentrate on constraints on $a_{\alpha\beta}^T$ by setting $c_{\alpha\beta}^{TT}=0$. The constraints on $c_{\alpha\beta}^{TT}$ will be commented later.

Recently, Super-Kamiokande \cite{Abe:2014wla} has set upper limits for $\vert a_{\alpha\beta}^T\vert$, which are of the order $10^{-23}$ GeV. With $\vert a_{\alpha\beta}^T\vert$ of this energy scale, it is interesting to note that $\Delta m^2_{ij}/E$ is smaller than 
$\vert a_{\alpha\beta}^T\vert $ by more than 3 orders of magnitude for neutrino energies beyond a few tens of TeV. Hence for neutrino events analyzed in IceCube flavor measurement~\cite{Aartsen:2015knd}, the LV term $H_{\rm LV}$ dominates over the standard model Hamiltonian
$UM^2U^{\dag}/2E $ if any $a_{\alpha\beta}^T$ term is set at the SK limit, $\sim10^{-23}~{\rm GeV}$. Therefore, IceCube measurements of flavor ratios should be useful for constraining the LV mass scale.

\begin{figure}[htbp]
	\begin{center}
	\includegraphics[width=10cm]{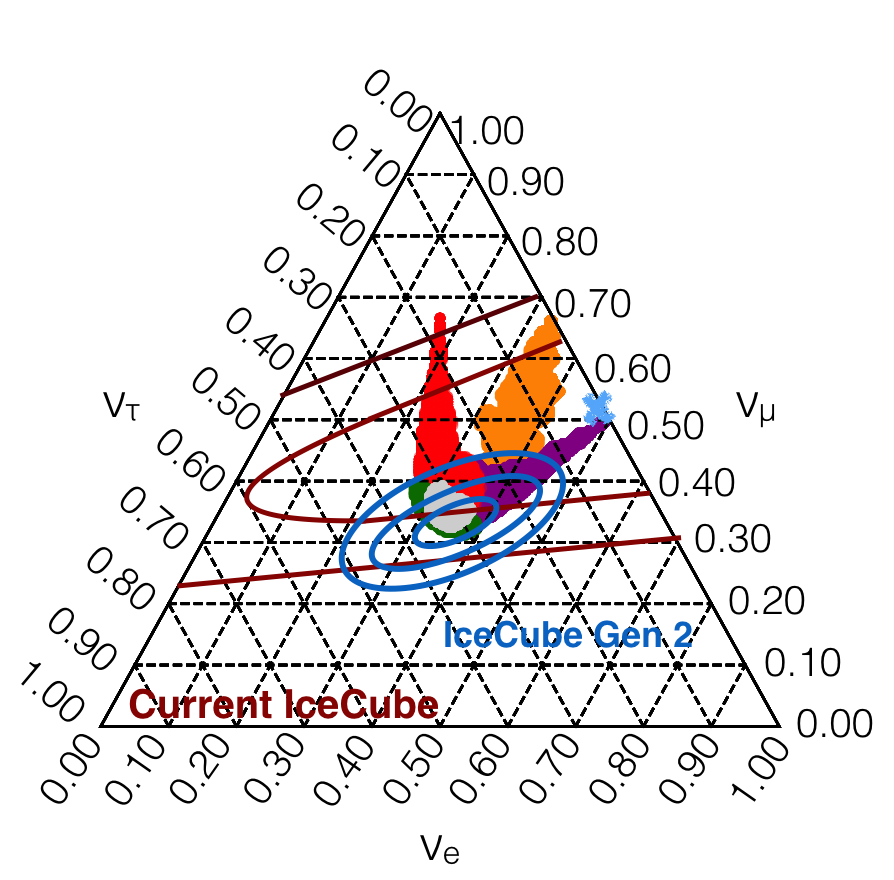}
	\caption{The flavor fractions of astrophysical neutrinos arriving on Earth. These neutrinos are assumed to come from the astrophysical pion source with 
	the flavor fraction $(1/3,~2/3,~0)$. The predicted ranges of flavor fractions on Earth by the full Hamiltonian $H^{\nu,\bar{\nu}}_{\rm SM}+H^{\nu,\bar{\nu}}_{\rm LV}$ are denoted by purple, red, gray, and orange areas for the special scenarios of $H^{\nu,\bar{\nu}}_{\rm LV}$ with $\vert a_{e\mu}^T\vert \neq 0$, $\vert a_{e\tau}^T\vert \neq 0$, $\vert a_{\mu\tau}^T\vert \neq 0$, and $a_{\mu\mu,\tau\tau}^T\neq 0$, respectively. In each scenario, the magnitude of the relevant matrix element $\vert a_{\alpha\beta}^T\vert$ is varied between $0$ and the current Super-Kamiokande limit. The green area is the accessible range of neutrino flavor fraction by $H^{\nu,\bar{\nu}}_{\rm SM}$ with neutrino mixing angles and CP phase varied over $3\sigma$ range. Regions inside the brown lines are the current  IceCube measurements with the blue cross denoting the best fit values~\cite{Aartsen:2015knd}. Regions inside the blue curves are the expected IceCube-Gen2 $1\sigma-3\sigma$ sensitivity regions given in~\cite{Shoemaker:2015qul}. }
	\label{fig:mscale}
	\end{center}
\end{figure}

To illustrate the current IceCube capability of constraining LV parameters, we calculate the accessible ranges of neutrino flavor fractions on Earth resulting from the full Hamiltonian 
$H^{\nu,\bar{\nu}}_{\rm SM}+H^{\nu,\bar{\nu}}_{\rm LV}$ and the astrophysical pion source for neutrinos with the flavor fraction $(1/3,~2/3,~0)$. For an illustrative purpose, we consider special scenarios for $H^{\nu,\bar{\nu}}_{\rm LV}$ where only one pair of 
matrix elements in LV Hamiltonian, for instance, $a_{\alpha\beta}^T$ and its complex conjugate $a_{\alpha\beta}^{T*}$, are non-vanishing.  We classify these special scenarios as $\vert a_{e\mu}^T\vert \neq 0$, $\vert a_{e\tau}^T\vert \neq 0$, $\vert a_{\mu\tau}^T\vert \neq 0$, and $a_{\mu\mu,\tau\tau}^T\neq 0$, respectively. For the last scenario we take $a_{\tau\tau}^T=-a_{\mu\mu}^T$. In each special scenario for $H^{\nu,\bar{\nu}}_{\rm LV}$,  the magnitude of the relevant matrix element $\vert a_{\alpha\beta}^T\vert $ is varied from zero to the current Super-Kamiokande $95\%$ C. L. limit, the phase of  $a_{\alpha\beta}^T$ is varied from $0$ to $2\pi$, and the neutrino mixing parameters in $H^{\nu,\bar{\nu}}_{\rm SM}$
are taken to be their best-fit values~\cite{Gonzalez-Garcia:2015qrr}. The predicted ranges of flavor fractions on Earth by the full Hamiltonian $H^{\nu,\bar{\nu}}_{\rm SM}+H^{\nu,\bar{\nu}}_{\rm LV}$ 
for all considered scenarios of LV Hamiltonian are shown in Fig.~\ref{fig:mscale}. We stress that $H^{\nu,\bar{\nu}}_{\rm LV}$ dictates the neutrino flavor fraction 
when $\vert a_{\alpha\beta}^T\vert$ is taken at the current SK limit in each special scenario. For comparison, the standard-model predicted neutrino flavor fractions with neutrino mixing angles and CP phase in $H^{\nu,\bar{\nu}}_{\rm SM}$ varied over $3\sigma$ range
\cite{Bustamante:2015waa} is also shown as the green area~\cite{comment_SM} in Fig.~\ref{fig:mscale}.  
It is clear that, except for a tiny piece of area, the predicted ranges of flavor fractions of neutrinos by the full Hamiltonian $H^{\nu,\bar{\nu}}_{\rm SM}+H^{\nu,\bar{\nu}}_{\rm LV}$ are all within the current IceCube $3\sigma$ contour. Therefore a stringent constraint to $H^{\nu,\bar{\nu}}_{\rm LV}$ requires IceCube-Gen2, which is the main target of our study in the next session.

\section{The sensitivity of IceCube-Gen2 to the LV parameters }

In this section, we apply the projected flavor discrimination sensitivity of IceCube-Gen2~\cite{Shoemaker:2015qul} to estimate the future constraints on LV
parameters. In the above projected sensitivity, only the pion source produced by $pp$ collisions is considered. Therefore we shall only consider this type of source in the following discussions.

Before studying constraints to the most general flavor structure of $H^{\nu,\bar{\nu}}_{\rm LV}$, it is useful to summarize our analysis in the 
previous section. Let us take $f_{\alpha}\equiv \Phi(\nu_\alpha)/(\Phi(\nu_e)+\Phi(\nu_\mu)+\Phi(\nu_\tau))$
as the neutrino flavor fraction on the Earth. Since we shall focus on the pion source caused by $pp$ collisions, there are equal numbers of
neutrinos and anti-neutrinos produced with the flavor fraction $(1/3,~2/3,~0)$ at the source for both neutrinos and anti-neutrinos. Therefore we have $f_{e}=P_{ee}/3+2P_{e\mu}/3$. Since $P_{\alpha\beta}=P_{\beta\alpha}$ still holds with the addition of LV Hamiltonian, we thus have $P_{ee}= 1-P_{\mu e}-P_{\tau e}
=1-P_{e\mu}-P_{e\tau}$. Hence $f_e=1/3+(P_{e \mu}-P_{e \tau})/3$. Similarly we can show that $f_{\mu}=1/3+(P_{\mu \mu}-P_{\mu \tau})/3$, and 
$f_{\tau}=1/3+(P_{\mu\tau}-P_{\tau \tau})/3$.  Clearly for astrophysical neutrinos arising from the pion source, the deviation of their flavor fraction on Earth to $(1/3,~1/3,~1/3)$ is due to $\mu-\tau$ symmetry breaking effects in 
the transition probability matrix. 
For the standard model Hamiltonian $H_{\rm SM}$, the $\mu-\tau$ symmetry breaking effects are small. To leading orders in $\cos2\theta_{23}$ and 
$\sin\theta_{13}$, one has $(P_{e \mu}-P_{e \tau})=2\epsilon$, $(P_{\mu\mu}-P_{\mu\tau})=(P_{\mu\tau}-P_{\tau\tau})=-\epsilon$ with
$\epsilon=2\cos2\theta_{23}/9+\sqrt{2}\sin\theta_{13}\cos\delta/9$ (taking $\sin^2\theta_{12}=1/3$)~\cite{Lai:2010tj} where $\delta$ is the CP violation phase. Hence LV effects can be detectable provided they introduce sizable $\mu-\tau$ symmetry breaking effects
in the neutrino flavor transition probability matrix. 

In the case that only $a^T_{e\mu}$ and $a^{T*}_{e\mu}$ are non-vanishing in $H^{\nu,\bar{\nu}}_{\rm LV}$, $\mu-\tau$ symmetry is clearly broken. If 
$H^{\nu,\bar{\nu}}_{\rm LV}$ dominates over $H^{\nu,\bar{\nu}}_{\rm SM}$, the flavor transition probability is determined by LV Hamiltonian and we find  
$(P_{e \mu}-P_{e \tau})=(P_{\mu\mu}-P_{\mu\tau})=1/2$ and $(P_{\mu\tau}-P_{\tau\tau})=-1$ in this limit. Consequently, the flavor fraction of astrophysical neutrinos arriving on Earth deviates significantly 
from $(1/3,~1/3,~1/3)$. This corresponds to the tip of purple area in Fig.~\ref{fig:mscale}, which represents the flavor fraction $(1/2,~1/2,~0)$.  Similarly, large $\mu-\tau$ symmetry breaking occurs in the scenarios $\vert a_{e\tau}^T\vert \neq 0$ and $a_{\mu\mu,\tau\tau}^T\neq 0$ ($a_{\mu\mu}^T\neq a_{\tau\tau}^T$).  
On the other hand, $\mu-\tau$ symmetry is preserved in the scenario $\vert a_{\mu\tau}^T\vert \neq 0$.

We have just seen that the $\mu-\tau$ symmetry breaking effect in $H^{\nu,\bar{\nu}}_{\rm LV}$ can be probed with the pion source produced by $pp$ collisions. 
Since we have assumed that all astrophysical neutrinos come from the pion source, it is essential to quantify the $\mu-\tau$ symmetry breaking effect in $H^{\nu,\bar{\nu}}_{\rm LV}$.  To do that, it is useful to write
$H^{\nu}_{\rm LV}=H^{\nu}_1+H^{\nu}_2$ with
\begin{equation}
H^{\nu}_1= \left(
                            \begin{array}{ccc}
                            0                 &     0             &  0 \\
                            0 &      a_{\mu\mu}^{T}                           & a_{\mu\tau}^{T } \\
                            0 &      a_{\mu\tau}^{T *}       & a_{\tau\tau}^{T}  
                            \end{array}
                            \right), \label{mu-tau-sector-a}
\end{equation}
and 
\begin{equation}
H^{\nu}_2= \left(
                            \begin{array}{ccc}
                            0                 &     a_{e\mu}^{T}             &  a_{e\tau}^{T} \\
                                a_{e\mu}^{T*}               &      0                           & 0 \\
                          a_{e\tau}^{T*}   &      0       & 0  
                            \end{array}
                            \right). \label{emutau-sector-a}
\end{equation}
Similar decomposition can be applied to $H^{\bar{\nu}}_{\rm LV}$. 

We note that the simplified structure $H^{\nu}_1$ has been considered as the LV coupling between dark energy and neutrinos
and the measurement of astrophysical $\nu_{\mu}$ and $\nu_{\tau}$ event difference was proposed to constrain $H^{\nu}_1$ in the future~\cite{Ando:2009ts}.   
Here we shall begin with simplified scenarios that $H^{\nu,\bar{\nu}}_{\rm LV}=H^{\nu,\bar{\nu}}_1$ and  $H^{\nu,\bar{\nu}}_{\rm LV}=H^{\nu,\bar{\nu}}_2$. We then proceed to discuss the general case with 
$H^{\nu,\bar{\nu}}_{\rm LV}=H^{\nu,\bar{\nu}}_1+H^{\nu,\bar{\nu}}_2$. We shall study the sensitivities of IceCube-Gen2 to these Hamiltonians. 
\subsection{  $H^{\nu,\bar{\nu}}_{\rm LV}=H^{\nu,\bar{\nu}}_1$}
For  $H^{\nu,\bar{\nu}}_{\rm LV}=H^{\nu,\bar{\nu}}_1$, we can write
\begin{equation}
H^{\nu}_1=\left(\frac{a_{\mu\mu}^T+a_{\tau\tau}^T}{2}\right) \left(
                            \begin{array}{ccc}
                            1                 &     0             &  0 \\
                            0 &      1                           & 0 \\
                            0 &      0      &  1  
                            \end{array}
                            \right)-\frac{1}{2}\left(
                            \begin{array}{ccc}
                            a_{\mu\mu}^T+a_{\tau\tau}^T                 &     0             &  0 \\
                            0 &      a_{\tau\tau}^T-a_{\mu\mu}^{T}                           &-2 a_{\mu\tau}^{T } \\
                            0 &      -2a_{\mu\tau}^{T *}       &  a_{\mu\mu}^T-a_{\tau\tau}^{T}  
                            \end{array}
                            \right).\label{mu-tau-sector-b}
\end{equation}
The first term of $H^{\nu}_1$ is proportional to the identity matrix and does not affect the neutrino flavor transition probability. One can ignore this term and rewrite $H^{\nu}_1$ as
\begin{equation}
H^{\nu}_1=-M\left(
                            \begin{array}{ccc}
                            \gamma                 &     0             &  0 \\
                            0 &      \cos 2\alpha                           &-e^{i\beta}\sin 2\alpha \\
                            0 &     -e^{-i\beta}\sin 2\alpha        &  -\cos 2\alpha
                            \end{array}
                            \right),\label{mu-tau-sector-c}
\end{equation}
where $M=\sqrt{(a_{\tau\tau}^T-a_{\mu\mu}^T)^2+4a_{\mu\tau}^Ta_{\mu\tau}^{T*}}/2$, 
$\gamma=(a_{\mu\mu}^T+a_{\tau\tau}^T)/\sqrt{(a_{\tau\tau}^T-a_{\mu\mu}^T)^2+4a_{\mu\tau}^Ta_{\mu\tau}^{T*}}$, 
$\cos 2\alpha=(a_{\tau\tau}^T-a_{\mu\mu}^T)/\sqrt{(a_{\tau\tau}^T-a_{\mu\mu}^T)^2+4a_{\mu\tau}^Ta_{\mu\tau}^{T*}}$, $\sin 2\alpha=
2\vert  a_{\mu\tau}^T\vert/\sqrt{(a_{\tau\tau}^T-a_{\mu\mu}^T)^2+4a_{\mu\tau}^Ta_{\mu\tau}^{T*}}$, and $\beta$ is the phase of $a_{\mu\tau}^T$. Since $\sin 2\alpha$ is positive by definition, $\alpha$ varies between $0$ and $\pi/2$. The Hamiltonian $H^{\bar{\nu}}_1$ can be inferred from
$H^{\nu}_1$ by the replacements $-M\to M$ and $\beta\to -\beta$. Taking into account the total Hamiltonian, $H^{\nu,\bar{\nu}}=H^{\nu,\bar{\nu}}_{\rm SM}+H_1^{\nu,\bar{\nu}}$, one can predict the neutrino flavor fraction on Earth assuming 
the initial neutrino flavor fraction at the source to be $(1/3,~2/3,~0)$.   We note that the neutrino energy appearing in $H^{\nu,\bar{\nu}}_{\rm SM}$ should in principle follow the $E^{-2.2}$ distribution with the threshold at $100$ TeV
according to Ref.~\cite{Shoemaker:2015qul}. However, for simplicity, we fix $E=100$ TeV. This is a conservative choice that makes $H^{\nu,\bar{\nu}}_{\rm SM}$ less suppressed in comparison to the
dominant $H_1^{\nu,\bar{\nu}}$.   

\begin{figure}[htbp]
	\begin{center}
	\includegraphics[width=12cm]{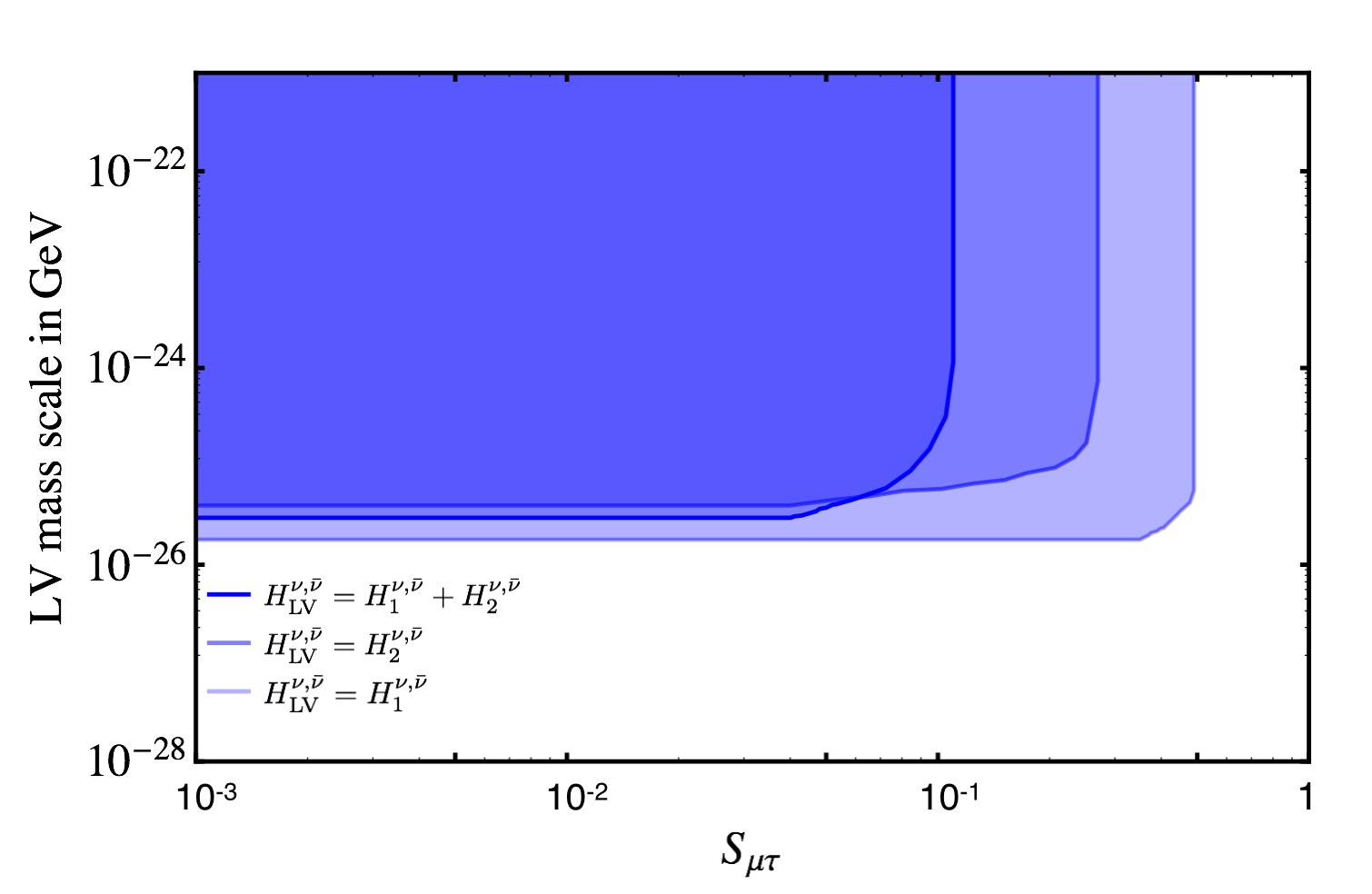}
	\caption{ The sensitivity of IceCube-Gen2 to the LV mass scale as a function of $\mu-\tau$ symmetry breaking parameter $S_{\mu\tau}$. The parameter range above each sensitivity curve will be ruled out at 
	$3\sigma$ if no deviation to the standard flavor transition of neutrinos is observed. These excluded ranges are obtained assuming the flavor fraction of astrophysical 
	neutrinos from each source is $(1/3,~2/3,~0)$ for all neutrino energies beyond $100$ TeV threshold. The LV mass scales for $H^{\nu,\bar{\nu}}_{\rm LV}= H^{\nu,\bar{\nu}}_1$ and 
	$H^{\nu,\bar{\nu}}_{\rm LV}= H^{\nu,\bar{\nu}}_2$ are $M$ and $M'$ defined in Eqs.~(\ref{mu-tau-sector-c}) and (\ref{emutau-sector-b}), respectively, while $S_{\mu\tau}$ for these two cases are $\sin^2 2\alpha$ and $\sin^2 2\rho$, respectively. The LV mass scale for $H^{\nu,\bar{\nu}}_{\rm LV}= H^{\nu,\bar{\nu}}_1+H^{\nu,\bar{\nu}}_2$ is $M$ under the assumption $M=M'$ and $S_{\mu\tau}$ for this case is $\sin 2\alpha \times \sin 2\rho$.  }
	\label{fig:constraint_dim3}
	\end{center}
\end{figure}

Given the IceCube-Gen2 sensitivity shown in Fig.~\ref{fig:mscale}, we obtain the expected constraints on the LV mass scale $M$
as a function of mixing angle $\alpha$ with the phase $\beta$ varied between $0$ and $2\pi$ and the ratio $\gamma$ of the order of unity. The expected constraints on $M$ are shown in that part of Fig.~\ref{fig:constraint_dim3} labeled by $H^{\nu,\bar{\nu}}_{\rm LV}= H^{\nu,\bar{\nu}}_1$. To derive the expected constraints on $M$, we first fix the $\mu-\tau$ symmetry breaking parameter $S_{\mu\tau}\equiv \sin^2 2\alpha$ while allow the parameters $\beta$ and
$\gamma$ to vary. We then identify the critical value of $M$ such that the resulting neutrino flavor fraction on the Earth reaches to the boundary of IceCube-Gen2 $3\sigma$ C.L. contour. 
In this way we obtain an expected constraint on $M$ for a specific $\sin^2 2\alpha$. We repeat the above procedure for different values of $\sin^2 2\alpha$ so that the entire sensitivity curve is obtained. The parameter range above the sensitivity curve will be ruled out at $3\sigma$ if no deviation to the standard neutrino flavor transition mechanism is observed.   

We note that the $\mu-\tau$ symmetry limit in $H^{\nu,\bar{\nu}}_1$ corresponds to $\sin^2 2\alpha=1$ while the maximum breaking
corresponds to $\sin^2 2\alpha=0$. This can be seen from the matrix structure given by Eq.~(\ref{mu-tau-sector-c}) or the neutrino flavor transition probabilities resulting from the Hamiltonian $H^{\nu,\bar{\nu}}_1$. 
For the latter we found $(P_{e \mu}-P_{e \tau})=0$, $(P_{\mu \mu}-P_{\mu \tau})=1-\sin^2 2\alpha$, and $(P_{\mu\tau}-P_{\tau \tau})=-1+\sin^2 2\alpha$. It is clear that $\sin^2 2\alpha$ indeed determines the above $\mu-\tau$ symmetry breaking effects in neutrino flavor transition probabilities.  For $0\leq \sin^2 2\alpha \leq 0.35$, the sensitivity of IceCube-Gen2 to $M$ is about $2\times 10^{-26} \ {\rm GeV}$. The sensitivity to $M$ diminishes for $\sin^2 2\alpha > 0.46$ ($\sin 2\alpha > 0.68$).
In our numerical studies, the neutrino mixing parameters 
in $H^{\nu,\bar{\nu}}_{\rm SM}$ are taken as the best fit values given in ~\cite{Gonzalez-Garcia:2015qrr}. This will be our choice for neutrino mixing parameters throughout the rest of the paper. We also vary each neutrino mixing parameter over $1\sigma$ range to see the effect. No appreciable 
effect in the sensitivity to $M$ is found.  
We note that the current SK $95\%$ C.L. limits on the related matrix elements are ${\rm Re}(a_{\mu\tau}^T)< 6.5\times 10^{-24}$ GeV and ${\rm Im}(a_{\mu\tau}^T)< 5.1\times 10^{-24}$ GeV~\cite{Abe:2014wla}. It is clear that the expected bounds by IceCube-Gen2 
shall improve the current bounds by more than two orders of magnitudes provided $\sin 2\alpha <0.68$. Particularly the IceCube Gen2 sensitivity presented here is at $3\sigma$ C.L.        
\subsection{  $H^{\nu,\bar{\nu}}_{\rm LV}=H^{\nu,\bar{\nu}}_2$}
For $H^{\nu,\bar{\nu}}_{\rm LV}=H^{\nu,\bar{\nu}}_2$, we can write
\begin{equation}
H^{\nu}_2=M^{\prime}\left(
                            \begin{array}{ccc}
                            0                 &     e^{i\sigma}\cos\rho             &  e^{i\lambda}\sin\rho \\
                             e^{-i\sigma}\cos\rho  &      0                           & 0 \\
                           e^{-i\lambda}\sin\rho  &     0        &  0
                            \end{array}
                            \right),\label{emutau-sector-b}
\end{equation}
where $M'=\sqrt{a_{e\mu}^T a_{e\mu}^{T*}+a_{e\tau}^T a_{e\tau}^{T*}}$, $\cos\rho=\vert a_{e\mu}^T\vert /M'$, $\sin\rho=\vert a_{e\tau}^T\vert /M'$, 
$\sigma$ and $\lambda$ are phases 
of $a_{e\mu}^T$ and $a_{e\tau}^T$, respectively. The Hamiltonian $H^{\bar{\nu}}_2$ can be inferred from
$H^{\nu}_2$ by the replacements $M^{\prime}\to -M^{\prime}$, $\sigma\to -\sigma$, and $\lambda\to -\lambda$. Since both $\cos\rho$ and $\sin\rho$ are positive by definition, 
the angle $\rho$ is between $0$ and $\pi/2$. Taking into account the total Hamiltonian, $H^{\nu,\bar{\nu}}=H^{\nu,\bar{\nu}}_{\rm SM}+H_2^{\nu,\bar{\nu}}$, one can predict the neutrino flavor fraction on Earth assuming 
the initial neutrino flavor fraction at the source is $(1/3,~2/3,~0)$. 

Given the IceCube-Gen2 sensitivity shown in Fig.~\ref{fig:mscale}, we obtain the expected constraints on the LV mass scale $M^{\prime}$
as a function of mixing angle $\rho$ with the phases $\sigma$ and $\lambda$ varied between $0$ and $2\pi$. The sensitivity to $M^{\prime}$ is shown in that part of Fig.~\ref{fig:constraint_dim3} labeled by $H^{\nu,\bar{\nu}}_{\rm LV}=H^{\nu,\bar{\nu}}_2$. 
We have varied each neutrino mixing parameter over $1\sigma$ range and no appreciable 
effect on the sensitivity to $M^{\prime}$ is found. 
The parameter $S_{\mu\tau}$ that characterizes the degree of $\mu-\tau$ symmetry breaking in $H^{\nu,\bar{\nu}}_2$ is $\sin^2 2\rho$. The $\mu-\tau$ symmetry limit corresponds to $\sin^2 2\rho=1$, i.e., $\rho=\pi/4$. On the other hand, the maximum breaking
corresponds to $\sin^2 2\rho=0$, i.e., $\rho=0$ or $\pi/2$. This is seen from the matrix structure given by Eq.~(\ref{emutau-sector-b}) or the neutrino flavor transition probabilities resulting from the Hamiltonian $H^{\nu,\bar{\nu}}_2$. For the latter one can show that the neutrino flavor transition probabilities depend on both $\sin 2\rho$ and $\cos 2\rho$.
Hence a specific value of $S_{\mu\tau}\equiv \sin^2 2\rho$ corresponds to two different neutrino flavor transition probabilities distinguished by the sign of $\cos 2\rho$. In principle there are two sensitivity points for each $S_{\mu\tau}$ but we have chosen the 
more conservative one to plot the sensitivity curve. 

For $0\leq \sin^2 2\rho \leq 0.2$, the sensitivity of IceCube-Gen2 to $M^{\prime}$ varies slowly from $4\times 10^{-26} \ {\rm GeV}$
to $7\times 10^{-26} \ {\rm GeV}$. In comparison, the current SK $95\%$ C.L. limits on related matrix elements are  ${\rm Re}(a_{e\mu}^T)< 1.8\times 10^{-23}$ GeV, ${\rm Im}(a_{e\mu}^T)< 1.8\times 10^{-23}$ GeV,
 ${\rm Re}(a_{e\tau}^T)< 4.1\times 10^{-23}$ GeV and ${\rm Im}(a_{e\tau}^T)< 2.8\times 10^{-23}$ GeV~\cite{Abe:2014wla}. One can see that the expected bounds by IceCube-Gen2 
shall improve the current bounds by more than two orders of magnitudes provided $\sin^2 2\rho \leq 0.2$.      
The sensitivity to $M^{\prime}$ diminishes for $\sin^2 2\rho > 0.27$ ($\sin 2\rho > 0.52$).  

\subsection{ $H^{\nu,\bar{\nu}}_{\rm LV}=H^{\nu,\bar{\nu}}_1+H^{\nu,\bar{\nu}}_2$}
For the general case with $H^{\nu,\bar{\nu}}_{\rm LV}=H^{\nu,\bar{\nu}}_1+H^{\nu,\bar{\nu}}_2$, the mass scales $M$ and $M^{\prime}$ of $H^{\nu,\bar{\nu}}_1$ and  $H^{\nu,\bar{\nu}}_2$, respectively, are independent parameters. These two scales can be comparable or one of the scales is suppressed in comparison to the other.  Since the latter scenario has already been discussed, we only focus on the former case. To simplify our discussions, we take $M=M^{\prime}$.  The sensitivity of IceCube-Gen2 to $M$ is shown in that part of  Fig.~\ref{fig:constraint_dim3} labeled by $H^{\nu,\bar{\nu}}_{\rm LV}=H^{\nu,\bar{\nu}}_1+H^{\nu,\bar{\nu}}_2$. The parameter $S_{\mu\tau}$ that characterizes the degree of $\mu-\tau$ symmetry breaking is $\sin 2\alpha \times \sin 2\rho$. For $\sin 2\alpha \times \sin 2\rho=1$, one must have both $\sin 2\alpha$ and $\sin 2\rho$ equal to unity, i.e., the $\mu-\tau$ symmetry is respected in both $H^{\nu,\bar{\nu}}_1$ and $H^{\nu,\bar{\nu}}_2$. For $\sin 2\alpha \times \sin 2\rho=0$, either $H^{\nu,\bar{\nu}}_1$ or $H^{\nu,\bar{\nu}}_2$ (or both) breaks $\mu-\tau$ symmetry maximally.
The sensitivity of IceCube-Gen2 to $M$ is $3\times 10^{-26}$ GeV for $0\leq \sin 2\alpha \times \sin 2\rho\leq 0.04$. The sensitivity becomes $10^{-25}$ GeV for $\sin 2\alpha \times \sin 2\rho= 0.08$. All these sensitivities improve significantly from the current SK bounds. 
The sensitivity of IceCube-Gen2 to $M$ diminishes for $\sin 2\alpha \times \sin 2\rho >0.11$. We also vary the neutrino mixing parameter in $1\sigma$ range and no appreciable effect on the sensitivity to $M$ is found. 

\subsection{Sensitivities to $c^{TT}_{\alpha\beta}$}
So far we have only discussed IceCube-Gen2 sensitivities to $a^T_{\alpha\beta}$. One can also study the sensitivities to parameters $c^{TT}_{\alpha\beta}$ by turning off $a^T_{\alpha\beta}$.  
Clearly  $-4Ec^{TT}_{\alpha\beta}/3$ replaces  $a^{T}_{\alpha\beta}$ when the latter is turned off. It should however be noted that, for the anti-neutrino case,  $c^{TT}_{\alpha\beta}$ is changed into $c^{TT*}_{\alpha\beta}$ while $a^T_{\alpha\beta}$ is turned into
$-a^{T*}_{\alpha\beta}$. 

\begin{figure}[htbp]
	\begin{center}
	\includegraphics[width=12cm]{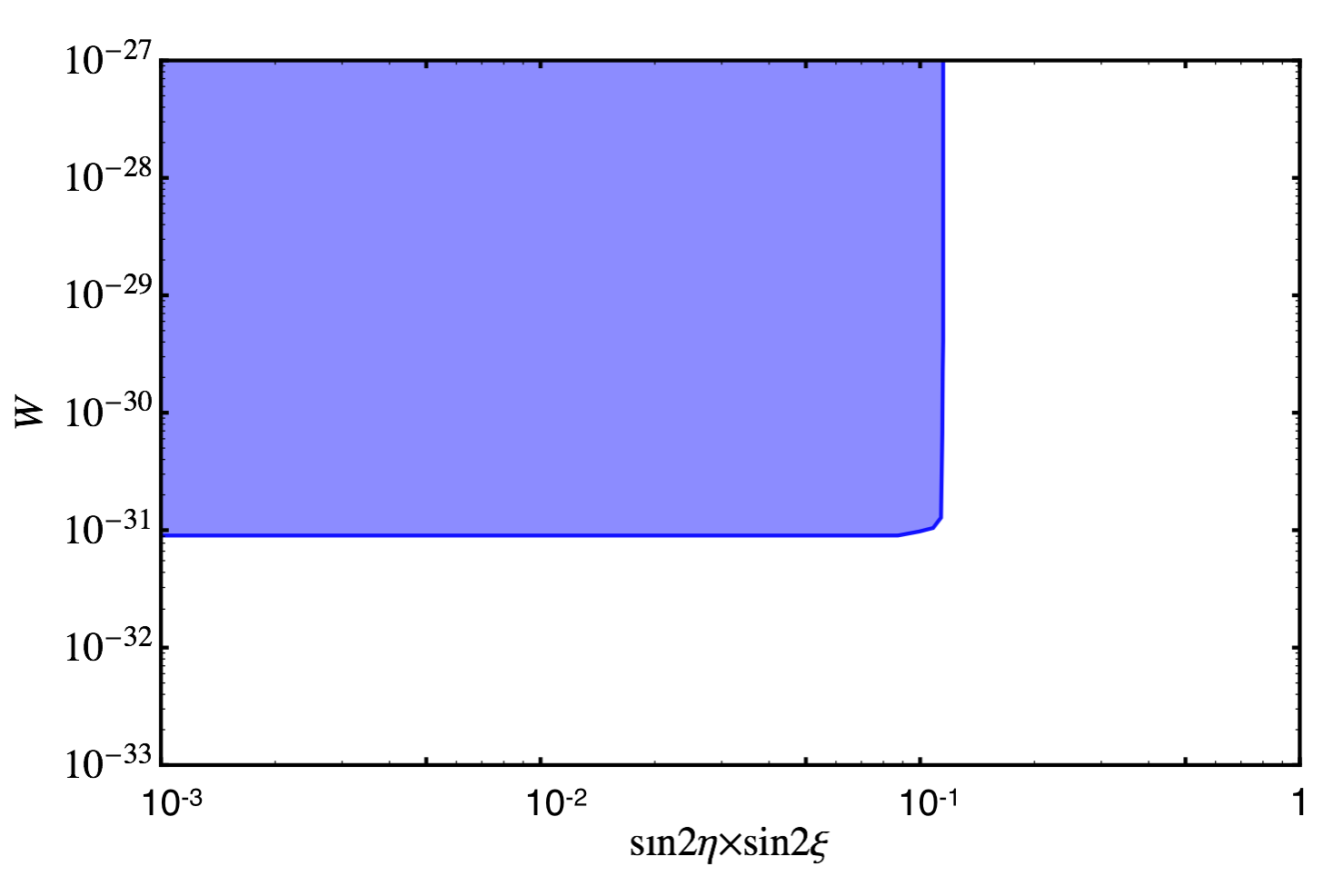}
	\caption{ The sensitivity of IceCube-Gen2 to the LV scale $W$ ($W=W^{\prime}$) of the Hamiltonian $\tilde{H}^{\nu,\bar{\nu}}_1+\tilde{H}^{\nu,\bar{\nu}}_2$ as a function of $\mu-\tau$ symmetry breaking parameter $\sin 2\eta \times  \sin 2\xi$. This sensitivity is obtained assuming the flavor fraction of astrophysical 
	neutrinos from each source is $(1/3,~2/3,~0)$ for all neutrino energies beyond $100$ TeV threshold. The parameter range above the sensitivity curve will be ruled out at 
	$3\sigma$ if no deviation to the standard flavor transition of neutrinos is observed.  }
	\label{fig:full_LV_dim4}
	\end{center}
\end{figure}

Following the previous treatment, one can also decompose the dimension-$4$, CPT-even LV Hamiltonian into two terms such that
\begin{equation}
\tilde{H}^{\nu}_1=-\frac{4E}{3} \left(
                            \begin{array}{ccc}
                            0                 &     0             &  0 \\
                            0 &      c_{\mu\mu}^{TT}                           & c_{\mu\tau}^{TT } \\
                            0 &      c_{\mu\tau}^{T T*}       & c_{\tau\tau}^{TT}  
                            \end{array}
                            \right), \label{mu-tau-sector-TT}
\end{equation} 
and 
\begin{equation}
\tilde{H}^{\nu}_2= -\frac{4E}{3}\left(
                            \begin{array}{ccc}
                            0                 &     c_{e\mu}^{TT}             &  c_{e\tau}^{TT} \\
                                c_{e\mu}^{TT*}               &      0                           & 0 \\
                          c_{e\tau}^{TT*}   &      0       & 0  
                            \end{array}
                            \right), \label{emutau-sector-TT}
\end{equation}
where we have used $\tilde{H}^{\nu,\bar{\nu}}_{1,2}$ to denote CPT-even LV Hamiltonian.
The LV Hamiltonian for anti-neutrinos can be obtained by taking complex conjugates. 
Analogous to our definitions of $M$ and $M^{\prime}$ from $a^{T}_{\alpha\beta}$, we can define dimensionless parameters  
$W\equiv \sqrt{(c_{\tau\tau}^{TT}-c_{\mu\mu}^{TT})^2+4c_{\mu\tau}^{TT}c_{\mu\tau}^{TT*}}/2$ and $W^{\prime}\equiv \sqrt{c_{e\mu}^{TT} c_{e\mu}^{TT*}+c_{e\tau}^{TT} c_{e\tau}^{TT*}}$, respectively. 
Let us consider the full LV Hamiltonian $H^{\nu,\bar{\nu}}_{\rm LV}=\tilde{H}^{\nu,\bar{\nu}}_1+\tilde{H}^{\nu,\bar{\nu}}_2$ and takes $W=W^{\prime}$. The sensitivity of IceCube-Gen2 to $W$ is shown in Fig.~\ref{fig:full_LV_dim4}. 
We have taken  $\sin 2\eta \times \sin 2\xi$ as the parameter to characterize the degree of  $\mu-\tau$ symmetry breaking, with $\sin 2\eta=
2\vert  c_{\mu\tau}^{TT}\vert/\sqrt{(c_{\tau\tau}^{TT}-c_{\mu\mu}^{TT})^2+4c_{\mu\tau}^{TT}c_{\mu\tau}^{TT*}}$ and $\sin\xi=\vert c_{e\tau}^{TT}\vert /W'$. Furthermore we also take $E=100$ TeV in $H^{\nu,\bar{\nu}}_{\rm LV}$ for simplicity. The sensitivity of IceCube-Gen2 to $W$ is about $10^{-31}$ for $0\leq \sin 2\eta \times \sin 2\xi\leq 0.12$. Such a sensitivity shall improve significantly from the current SK $95\%$ C.L. limits,  ${\rm Re}(c_{\mu\tau}^{TT})< 4.4\times 10^{-27}$, ${\rm Im}(c_{\mu\tau}^{TT})< 4.2\times 10^{-27}$,   ${\rm Re}(c_{e\mu}^{TT})< 8.0\times 10^{-27}$, ${\rm Im}(c_{e\mu}^{TT})< 8.0\times 10^{-27}$, and much less stringent constraints on $c_{e\tau}^{TT}$. 
The sensitivity curve rises up immediately for $\sin 2\eta \times \sin 2\xi> 0.12$. This behavior is quite distinct from the behavior of sensitivity curve in Fig.~\ref{fig:constraint_dim3} which rises mildly in the range $0.04\leq \sin 2\alpha \times  \sin 2\rho\leq 0.08$ before its sharp rise 
at  $\sin 2\alpha \times  \sin 2\rho=0.11$. We attribute the shape difference between two sensitivity curves to the sign difference between $a_{\alpha\beta}^T$ and $c_{\alpha\beta}^{TT}$ terms.   To see this we change the sign of  
$c_{\alpha\beta}^{TT}$ ($c_{\alpha\beta}^{TT*}$) terms in the neutrino sector while keeping the sign of  $c_{\alpha\beta}^{TT*}$ ($c_{\alpha\beta}^{TT}$) in the anti-neutrino sector unchanged. It is found that the shape of sensitivity curve
in Fig.~\ref{fig:full_LV_dim4} is completely identical to the shape of sensitivity curve in  Fig.~\ref{fig:constraint_dim3} as it should be according to Eqs.~(\ref{hamilton_nu}) and (\ref{hamilton_nubar}).

\section{Discussions and Conclusions}  
In this paper, we discuss the sensitivities of future IceCube-Gen2 to Lorentz violation parameters in the neutrino sector. We consider the effects of Lorentz violating Hamiltonian on the flavor transitions of astrophysical neutrinos coming from the pion source produced by $pp$ collisions. In such a case, there are equal numbers of
neutrinos and anti-neutrinos produced with the flavor fraction $(1/3,~2/3,~0)$ at the source for both neutrinos and anti-neutrinos.  We have shown that the flavor fraction of such neutrinos as they arrive at the Earth is $(1/3,~1/3,~1/3)$ if
the neutrino Hamiltonian respects $\mu-\tau$ symmetry. The deviation to such a flavor fraction is therefore controlled by the breaking of $\mu-\tau$ symmetry in the neutrino Hamiltonian. For both CPT-odd and CPT-even LV Hamiltonian, we
decompose the LV Hamiltonian into two matrix structures as shown in Eqs.~(\ref{mu-tau-sector-a}), (\ref{emutau-sector-a}), (\ref{mu-tau-sector-TT}), and (\ref{emutau-sector-TT}). For each matrix structure we define the parameter that characterizes the degree of $\mu-\tau$ symmetry breaking and the scale of the matrix to be probed by the measurement of astrophysical neutrino flavor fractions.

Since the neutrino Hamiltonian in the Standard Model is approximately $\mu-\tau$ symmetric, the effect from the new physics Hamiltonian is important only when this Hamiltonian significantly breaks the $\mu-\tau$ symmetry.
Taking Fig.~\ref{fig:constraint_dim3} as an example, the LV Hamiltonian $H^{\nu,\bar{\nu}}_1$  breaks the $\mu-\tau$ symmetry significantly for $\sin^2 2\alpha\leq 0.46$ ($\sin 2\alpha \leq 0.68$) such that
the expected constraint to the LV mass scale $M$ by IceCube-Gen2 is stringent. It is of interest to see how restricted the parameter range  $0\leq \sin 2\alpha \leq 0.68$ is. Without specific preference to the  detailed structure of 
 $H^{\nu,\bar{\nu}}_1$, one can assume the angle $\alpha$ to be uniformly distributed from $0$ to $\pi/2$ for a fixed LV mass scale $M$. The condition $0\leq \sin 2\alpha \leq 0.68$ requires either $0\leq 2\alpha \leq 0.75$ or $\pi-0.75 \leq 2\alpha\leq \pi$. Such a range for $\alpha$
occupies $1.5/\pi\equiv 48\%$ of the total parameter space for $\alpha$. For $H^{\nu,\bar{\nu}}_2$, the LV mass scale $M^{\prime}$ is testable for the parameter range $0\leq \sin 2\rho \leq 0.52$. Assuming $\rho$ is uniformly distributed 
 between $0$ and $\pi/2$, the range for $\rho$ required by the above condition occupies about $35\%$ of the total parameter space for $\rho$.  Finally for the case of full LV Hamiltonian with $M=M^{\prime}$, 
 the LV mass scale $M$ is testable in the parameter range $\sin 2\alpha \times \sin 2\rho \leq 0.11$. This is $21\%$ of the total parameter space of $\alpha$ and $\rho$ evaluated by a simple Monte Carlo. In the case of CPT-even LV Hamiltonian,
 the dimensionless LV scale $W$ ($W=W^{\prime}$) of  $\tilde{H}^{\nu,\bar{\nu}}_{1}+\tilde{H}^{\nu,\bar{\nu}}_{2}$ is testable for $\sin 2\eta \times \sin 2\xi \leq 0.12$. Clearly the percentage of total parameter space of $\eta$ and $\xi$ that satisfies this condition is also around $20\%$. 
 
In summary, we have taken a phenomenological approach that incorporate all LV effects in the neutrino sector with a set of local operators~\cite{Colladay:1996iz,Colladay:1998fq,Kostelecky:2003fs,AmelinoCamelia:2005qa,Bluhm:2005uj}.  
We only focus on the isotropic LV effects~\cite{ Kostelecky:2003cr} so that the structure of LV Hamiltonian is given by Eqs.~(\ref{hamilton_nu}) and (\ref{hamilton_nubar}). We have worked out the sensitivities of IceCube-Gen2 to CPT-odd LV parameter $a^T_{\alpha\beta}$ originated from dimension-$3$ operator  and the CPT-even LV parameters $c^{TT}_{\alpha\beta}$ originated from dimension-$4$ operators.
We have shown that the expected IceCube-Gen2 sensitivities to LV mass scales can improve the current SK bounds~\cite{Abe:2014wla} by at least 
two orders of magnitudes for sufficiently large $\mu-\tau$ symmetry breaking effects in LV Hamiltonian. We reiterate again that our results are based upon the assumption that    
 all sources of astrophysical neutrinos have an energy independent flavor fraction for neutrinos at $(1/3,~2/3,~0)$. It is worthwhile to pursue further studies with both the energy
 dependence of neutrino flavor fraction and the variations of neutrino flavor fractions among different sources taken into account.

\section*{Acknowledgements}
We thank M.~Bustamante for useful comments. This work is supported by Ministry of Science and Technology, Taiwan  under Grant Nos. 106-2112-M-182-001 and 105-2112-M-009 -014.

{\it Note added.}---As we were revising this paper, we became aware of the newest IceCube analysis on Lorentz violation effects in neutrino sector using atmospheric
neutrino data~\cite{Aartsen:2017ibm}, which sets $99\%$ C.L. bounds on 
${\rm Re}(a_{\mu\tau}^T)$ and ${\rm Im}(a_{\mu\tau}^T)$ at $2.9\times 10^{-24}$ GeV and $99\%$ C.L. bounds on ${\rm Re}(c_{\mu\tau}^{TT})$ and ${\rm Im}(c_{\mu\tau}^{TT})$
at $3.9\times 10^{-28}$.


\begin{thebibliography}{99}

\bibitem{Kostelecky:1988zi} 
 V.~A.~Kostelecky and S.~Samuel,
 Phys.\ Rev.\ D {\bf 39}, 683 (1989).

\bibitem{Kostelecky:1991ak} 
  V.~A.~Kostelecky and R.~Potting,
  Nucl.\ Phys.\ B {\bf 359}, 545 (1991).

\bibitem{Colladay:1996iz} 
  D.~Colladay and V.~A.~Kostelecky,
  Phys.\ Rev.\ D {\bf 55}, 6760 (1997).

\bibitem{Colladay:1998fq} 
  D.~Colladay and V.~A.~Kostelecky,
  Phys.\ Rev.\ D {\bf 58}, 116002 (1998).

\bibitem{Kostelecky:2003fs} 
  V.~A.~Kostelecky,
  Phys.\ Rev.\ D {\bf 69}, 105009 (2004).
  
\bibitem{AmelinoCamelia:2005qa} 
  G.~Amelino-Camelia, C.~Lammerzahl, A.~Macias and H.~Muller,
  AIP Conf.\ Proc.\  {\bf 758}, 30 (2005).
  
\bibitem{Bluhm:2005uj} 
  R.~Bluhm,
  Lect.\ Notes Phys.\  {\bf 702}, 191 (2006).

\bibitem{Kostelecky:2008ts} 
  V.~A.~Kostelecky and N.~Russell,
  Rev.\ Mod.\ Phys.\  {\bf 83}, 11 (2011).
  
\bibitem{Mattingly:2005re} 
  D.~Mattingly,
  Living Rev.\ Rel.\  {\bf 8}, 5 (2005).
 
 \bibitem{Kostelecky:2003cr} 
  V.~A.~Kostelecky and M.~Mewes, Phys.\ Rev.\ D {\bf 69}, 016005 (2004). 
 
 \bibitem{Kostelecky:2003xn} 
  V.~A.~Kostelecky and M.~Mewes,
  Phys.\ Rev.\ D {\bf 70}, 031902 (2004).
  
\bibitem{Kostelecky:2004hg} 
  V.~A.~Kostelecky and M.~Mewes,
  Phys.\ Rev.\ D {\bf 70}, 076002 (2004).
  
\bibitem{Auerbach:2005tq} 
  L.~B.~Auerbach {\it et al.} [LSND Collaboration], Phys.\ Rev.\ D {\bf 72}, 076004 (2005).

\bibitem{Adamson:2008aa} 
  P.~Adamson {\it et al.} [MINOS Collaboration], Phys.\ Rev.\ Lett.\  {\bf 101}, 151601 (2008).

\bibitem{AguilarArevalo:2011yi} 
  A.~A.~Aguilar-Arevalo {\it et al.} [MiniBooNE Collaboration], Phys.\ Lett.\ B {\bf 718}, 1303 (2013).

\bibitem{Adamson:2012hp} 
  P.~Adamson {\it et al.} [MINOS Collaboration], Phys.\ Rev.\ D {\bf 85}, 031101 (2012).

\bibitem{Adamson:2010rn} 
  P.~Adamson {\it et al.} [MINOS Collaboration], Phys.\ Rev.\ Lett.\  {\bf 105}, 151601 (2010).

\bibitem{Rebel:2013vc} 
  B.~Rebel and S.~Mufson, Astropart.\ Phys.\  {\bf 48}, 78 (2013).

\bibitem{Abe:2012gw} 
  Y.~Abe {\it et al.} [Double Chooz Collaboration], Phys.\ Rev.\ D {\bf 86}, 112009 (2012).

\bibitem{Diaz:2013iba} 
  J.~S.~Diaz, T.~Katori, J.~Spitz and J.~M.~Conrad, Phys.\ Lett.\ B {\bf 727}, 412 (2013).

\bibitem{Abbasi:2010kx} 
  R.~Abbasi {\it et al.} [IceCube Collaboration], Phys.\ Rev.\ D {\bf 82}, 112003 (2010).

\bibitem{Abe:2014wla} 
  K.~Abe {\it et al.} [Super-Kamiokande Collaboration], Phys.\ Rev.\ D {\bf 91}, no. 5, 052003 (2015).

\bibitem{dispersion}
We note that the neutrino dispersion relation can be modified by Lorentz violation. Thus  it is possible that neutrino loses energy via Cherenkov radiation during its propagation. 
For astrophysical neutrinos, such an energy loss mechanism can also lead to stringent bounds on Lorentz violation as pointed out in J.~S.~Diaz, A.~Kostelecky and M.~Mewes,
  Phys.\ Rev.\ D {\bf 89}, no. 4, 043005 (2014).  

  \bibitem{Aartsen:2013bka} 
  M.~G.~Aartsen {\it et al.} [IceCube Collaboration], Phys.\ Rev.\ Lett.\  {\bf 111}, 021103 (2013).

\bibitem{Aartsen:2013jdh} 
  M.~G.~Aartsen {\it et al.} [IceCube Collaboration], Science {\bf 342}, 1242856 (2013).

\bibitem{Aartsen:2013eka} 
  M.~G.~Aartsen {\it et al.} [IceCube Collaboration], Phys.\ Rev.\ D {\bf 89}, no. 6, 062007 (2014).

\bibitem{Aartsen:2014gkd} 
  M.~G.~Aartsen {\it et al.} [IceCube Collaboration], Phys.\ Rev.\ Lett.\  {\bf 113}, 101101 (2014).

\bibitem{Aartsen:2015ivb} 
  M.~G.~Aartsen {\it et al.} [IceCube Collaboration], Phys.\ Rev.\ Lett.\  {\bf 114}, no. 17, 171102 (2015).

\bibitem{Aartsen:2015knd} 
  M.~G.~Aartsen {\it et al.} [IceCube Collaboration], Astrophys.\ J.\  {\bf 809}, no. 1, 98 (2015).

\bibitem{Mena:2014sja} 
  O.~Mena, S.~Palomares-Ruiz and A.~C.~Vincent, Phys.\ Rev.\ Lett.\  {\bf 113}, 091103 (2014).


\bibitem{Winter:2014pya} 
  W.~Winter, Phys.\ Rev.\ D {\bf 90}, no. 10, 103003 (2014).

\bibitem{Chen:2014gxa} 
C.~Y.~Chen, P.~S.~Bhupal Dev and A.~Soni,
 Phys.\ Rev.\ D {\bf 92}, no. 7, 073001 (2015).

\bibitem{Palomares-Ruiz:2015mka} 
  S.~Palomares-Ruiz, A.~C.~Vincent and O.~Mena, Phys.\ Rev.\ D {\bf 91}, no. 10, 103008 (2015).

\bibitem{Palladino:2015zua} 
  A.~Palladino, G.~Pagliaroli, F.~L.~Villante and F.~Vissani, Phys.\ Rev.\ Lett.\  {\bf 114}, no. 17, 171101 (2015).

\bibitem{Aartsen:2014njl} 
  M.~G.~Aartsen {\it et al.} [IceCube Collaboration],
arXiv:1412.5106 [astro-ph.HE].

\bibitem{Aartsen:2015dkp} 
  M.~G.~Aartsen {\it et al.} [IceCube Collaboration],
  arXiv:1510.05228 [astro-ph.IM].
 %
 
 \bibitem{Shoemaker:2015qul} 
  I.~M.~Shoemaker and K.~Murase,
  Phys.\ Rev.\ D {\bf 93}, no. 8, 085004 (2016).

\bibitem{Lipari:2007su} 
  P.~Lipari, M.~Lusignoli and D.~Meloni, Phys.\ Rev.\ D {\bf 75}, 123005 (2007).


\bibitem{Kashti:2005qa} T. Kashti and E. Waxman, Phys. Rev. Lett. {\bf 95}, 181101(2005).

\bibitem{Kachelriess:2007tr} 
  M.~Kachelriess, S.~Ostapchenko and R.~Tomas, Phys.\ Rev.\ D {\bf 77}, 023007 (2008).
  
\bibitem{Hummer:2010ai} 
  S.~Hummer, M.~Maltoni, W.~Winter and C.~Yaguna, Astropart.\ Phys.\  {\bf 34}, 205 (2010).
  
\bibitem{Murase:2005hy} 
K.~Murase and S.~Nagataki,
Phys.\ Rev.\ D {\bf 73}, 063002 (2006).

\bibitem{Baerwald:2010fk} 
 P.~Baerwald, S.~Hummer and W.~Winter,
Phys.\ Rev.\ D {\bf 83}, 067303 (2011).


  

  
\bibitem{Arguelles:2015dca} 
  C.~A.~Arg\"{u}elles, T.~Katori and J.~Salvado,
  Phys.\ Rev.\ Lett.\  {\bf 115}, 161303 (2015).
  
 \bibitem{Katori:2016eni} 
  T.~Katori, C.~A.~Arg\"{u}elles and J.~Salvado,
  arXiv:1607.08448 [hep-ph]. 

\bibitem{Barenboim:2003jm} 
  G.~Barenboim and C.~Quigg,
  Phys.\ Rev.\ D {\bf 67}, 073024 (2003).
  
\bibitem{Hooper:2005jp} 
  D.~Hooper, D.~Morgan and E.~Winstanley,
  Phys.\ Rev.\ D {\bf 72}, 065009 (2005).
  
\bibitem{Bustamante:2010nq} 
  M.~Bustamante, A.~M.~Gago and C.~Pena-Garay,
  JHEP {\bf 1004}, 066 (2010).

\bibitem{Bustamante:2015waa} 
  M.~Bustamante, J.~F.~Beacom and W.~Winter,
  Phys.\ Rev.\ Lett.\  {\bf 115}, no. 16, 161302 (2015).
  
\bibitem{Gonzalez-Garcia:2015qrr} 
  M.~C.~Gonzalez-Garcia, M.~Maltoni and T.~Schwetz,
  Nucl.\ Phys.\ B {\bf 908}, 199 (2016).
\bibitem{comment_SM}
Here we recalculate the range of neutrino flavor fractions by $H^{\nu,\bar{\nu}}_{\rm SM}$ with updated fitting results for neutrino mixing angles and CP phase given in Ref.~\cite{Gonzalez-Garcia:2015qrr}.
  
\bibitem{Lai:2010tj} 
 K.~C.~Lai, G.~L.~Lin and T.~C.~Liu,
 Phys.\ Rev.\ D {\bf 82}, 103003 (2010).

\bibitem{Ando:2009ts} 
  S.~Ando, M.~Kamionkowski and I.~Mocioiu,
  Phys.\ Rev.\ D {\bf 80}, 123522 (2009)
  doi:10.1103/PhysRevD.80.123522
  [arXiv:0910.4391 [hep-ph]].


\bibitem{Aartsen:2017ibm} 
  M.~G.~Aartsen {\it et al.} [IceCube Collaboration],
  arXiv:1709.03434 [hep-ex].
  




\end{thebibliography}
\end{document}